\documentclass{PoS}

\usepackage{amsmath, multirow, enumitem}

\usepackage{tikz}
\usetikzlibrary{shapes.geometric, arrows}

\usepackage[natbib, sorting=none, maxcitenames=3, maxbibnames=3, style=numeric, backend=bibtex]{biblatex}

\DeclareFieldFormat*{title}{\emph{#1}}

\title{A citizen-science approach to muon events in imaging atmospheric Cherenkov telescope data: the Muon Hunter}

\ShortTitle{A citizen-science approach to muon events in IACT data: the Muon Hunter}

\tikzstyle{startstop} = [rectangle, rounded corners, minimum width=3cm, minimum height=1cm,text centered, draw=black, fill=red!20]
\tikzstyle{io} = [trapezium, trapezium left angle=70, trapezium right angle=110, minimum width=3cm, minimum height=1cm, text centered, draw=black, fill=blue!20]
\tikzstyle{process} = [rectangle, minimum width=3cm, minimum height=1cm, text centered, draw=black, fill=orange!20]
\tikzstyle{decision} = [diamond, minimum width=3cm, minimum height=1cm, text centered, draw=black, fill=green!20]
\tikzstyle{arrow} = [thick,->,>=stealth]

\author{\speaker{Q.~Feng}$^a$ for the VERITAS Collaboration\thanks{veritas.sao.arizona.edu}, J.~Jarvis$^{b}$ \\
         \llap{$^a$}Physics Department, McGill University, Montreal, QC H3A 2T8, Canada\\
         E-mail: \email{qi.feng2@mcgill.ca} \\
\llap{$^b$} ASTERICS, DECS Workpackage, School of Physical Sciences, The Open University, Milton Keynes MK7 6AA, UK \\
}

\abstract{
Event classification is a common task in gamma-ray astrophysics. It can be treated with rapidly-advancing machine learning algorithms, which have the potential to outperform traditional analysis methods. However, a major challenge for machine learning models is extracting reliably labelled training examples from real data. Citizen science offers a promising approach to tackle this challenge. 

We present "Muon Hunter", a citizen science project hosted on the Zooniverse platform, where VERITAS data are classified multiple times by individual users in order to select and parameterize muon events, a product from cosmic ray induced showers. We use this dataset to train and validate a convolutional neural-network model to identify muon events for use in monitoring and calibration. The results of this work and our experience of using the Zooniverse are presented. 
}

\FullConference{35th International Cosmic Ray Conference - ICRC2017\\
		12-20 July, 2017\\
		Bexco, Busan, Korea}

\begin{document}

\section{Introduction} \label{sec:intro}

In the past decade, our understanding of the very-high-energy (VHE; 100~GeV $\lesssim E_{\gamma} \lesssim$ 100~TeV) gamma-ray sky has greatly progressed through the use of imaging atmospheric Cherenkov telescopes (IACTs). IACTs image the Cherenkov light of an extensive air shower induced by an incident VHE gamma-ray photon or a cosmic-ray (CR) particle. The air-shower images are then analyzed to reconstruct the information of the incident photons or the CR particles, of which the latter comprise a substantial background in VHE gamma-ray astronomy. 
The ability to separate gamma rays from CR particles is important, as it is directly related to the sensitivity of the instrument. 
Such separation is performed based on the morphology of the shower images, and can be formalized as a computer vision classification task. 

Classification is a common task in experimental physics. Usually, a few key parameters are chosen, and cuts determined from Monte Carlo simulations are applied to the chosen parameters~\citep[e.g.][]{Hillas85}. More recently, the use of machine learning algorithms has become increasingly popular in many branches of physics, including VHE gamma-ray astrophysics to improve event classification~\citep[e.g.][]{Albert08, Acero09, Krause17, Feng17}. 
One powerful machine learning algorithm, convolutional neural networks (CNN)~\citep{Krizhevsky12}, was used on a small batch of VHE gamma-ray data to detect and characterize muon events~\cite{Feng17}. 
Muons are secondary products in CR-particle air showers that propagate to the ground and emit Cherenkov light, detected as rings or arcs by IACTs. 
The number of Cherenkov photons from a muon reaching the reflector of a telescope can be estimated from the Cherenkov angle and the traveling direction of the muon, and the distance between the telescope and where the muon impact the ground. 
Therefore, muon events can be used as a calibration source for the throughput of IACTs~\citep[e.g.][]{Hanna08, Tyler13}. 
%

For developing supervised machine learning algorithms like the CNN, 
it is essential to provide correct labels of a large training dataset. This was done using a standard analysis in~\cite{Feng17}. 
In this work, we describe the Muon Hunter\footnote{\url{www.muonhunter.org}}, a citizen-science project where volunteers label and parameterize muon and non-muon images in VHE gamma-ray data. 

\section{The VERITAS array and a standard muon analysis}
\label{sec:V}

The Very Energetic Radiation Imaging Telescope Array System (VERITAS) is an array of
four IACTs located at the Fred Lawrence Whipple Observatory in southern Arizona \citep[30$^\circ$40'N 110$^\circ$57'W, 1.3 km a.s.l.;][]{Holder11}. It is sensitive to gamma rays in the energy range from 85 GeV to $>$30 TeV with an energy resolution of $\sim$15\% (at 1~TeV). 
Each of the four telescopes is equipped with a 12-m diameter reflector comprising 345 identical mirror facets, 
and a 499-pixel photomultiplier tube (PMT) camera covering a field of view of 3.5$^\circ$. The array has a 68\%-containment gamma-ray angular resolution at 1~TeV of $\sim$0.1$^{\circ}$. 

Coincident 
signals from at least two out of the four telescopes are required to trigger an array-wide read-out of the PMT signals, which occurs at a typical rate of $\sim$400 Hz. 
Most of these triggers come from CR showers or night sky background noise. For comparison, the brightest steady VHE source, the Crab Nebula, is typically observed by VERITAS at a rate of $<$15 gamma rays per minute. 
Muons are produced from charged pion decays in CR hadronic showers and are background for gamma-ray astrophysics. Single muons reaching ground level usually only trigger one telescope and produce ring images. 
Therefore the multi-telescope trigger system greatly reduces the muon trigger rate. 
There are still muon events in the data because the array can be triggered by sub-showers or energetic muons impacting the ground between two telescopes. 

We randomly selected 16 observations, each of which is 30 minutes in duration, and analyzed them using one of the standard VERITAS data analysis packages, named \textit{VEGAS} \citep{Cogan08}. 
The procedure for the \textit{VEGAS} analysis of muon events is described as follows \cite[see also][]{Feng17}: 
\begin{enumerate}[label=(\arabic*)]
\item calculate the brightness-weighted average coordinates of the image (see e.g. Figure~\ref{fig:rings}), and use them as the initial muon-ring center; 
\item calculate the mean ($\bar{r}$) and the variance ($\sigma^2_r$) of the distances between all image pixels and the initial centroid, and then use $\bar{r}$ as the initial muon radius; 
\item move the initial centroid by a small step, repeat step (2) and check if the variance $\sigma^2_r$ decreases; if so, update the centroid, $\bar{r}$ and the variance $\sigma^2_r$; 
\item repeat step (3) to cover a predefined grid around the initial centroid, and return the optimal centroid and radius that minimize $\sigma^2_r$; 
\item check if $>$70\% of the pixels fall into a predefined accepted annulus e.g. $(\bar{r}-1.5 \sigma_r, \bar{r} +1.5 \sigma_r)$; if so, accept this event as a muon event. 
\end{enumerate}

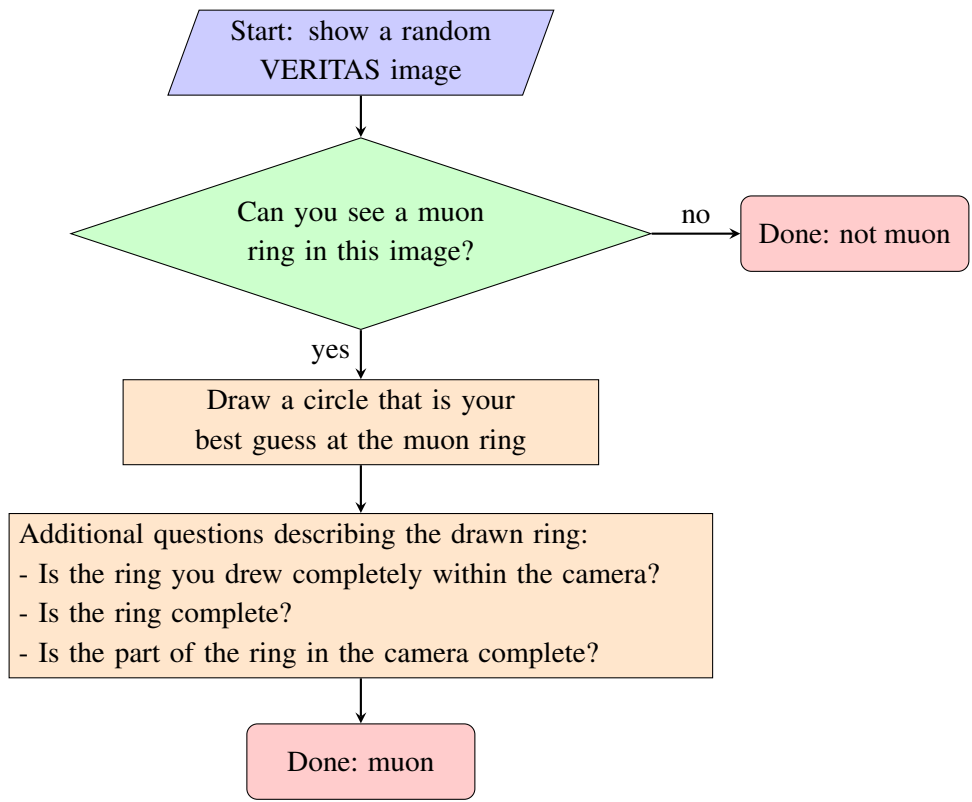
\begin{figure}[b]
 \centering 
\begin{tikzpicture}[node distance=2cm]
\node (in1) [io, text width=4cm] {Start: show a random VERITAS image};
\node (dec1) [decision, below of=in1, aspect=3, text width=4cm, yshift=-0.4cm] {Can you see a muon ring in this image?};

\node (proc1) [process, below of=dec1, yshift=-0.5cm, text width=6cm] {Draw a circle that is your best guess at the muon ring};
\node (proc2) [process, below of=proc1, yshift=-0.3cm, text width=9cm, align=left] {Additional questions describing the drawn ring: \\ 
															- Is the ring you drew completely within the camera? \\ 
															- Is the ring complete? \\
															- Is the part of the ring in the camera complete?};

\node (stop2) [startstop, right of=dec1, xshift=4.5cm] {Done: not muon};

\node (stop) [startstop, below of=proc2, yshift=-0.2cm] {Done: muon};

\draw [arrow] (in1) -- (dec1);
\draw [arrow] (dec1) -- node[anchor=east] {yes} (proc1);
\draw [arrow] (dec1) -- node[anchor=south] {no} (stop2);
\draw [arrow] (proc1) -- (proc2);
\draw [arrow] (proc2) -- (stop);

\end{tikzpicture}
\caption{An illustration of the workflow of Muon Hunter. 
\label{fig:flowchart}}
\end{figure}
A double-pass method based on the above procedure was used in the analysis, 
and dead or stuck pixels were corrected for. 
We used the analysis described above as one way to label muon events for the training/test data to build a CNN model, with an additional requirement that $\bar{r} \geqslant 0.5^\circ$. 
We note that an independent muon analysis using the Hough transform \cite{Tyler13} is available but was not used in this work. 
\label{sec:MH}
%

\section{The Muon Hunter project}

The Muon Hunter experiment was developed in collaboration with the ASTERICS Horizon2020 project\footnote{\url{www.asterics2020.eu/}}, and the website is hosted by the Zooniverse\footnote{\url{www.zooniverse.org}} platform, where researchers in many disciplines can easily publish a citizen-science project through the Zooniverse Project Builder\footnote{\url{www.zooniverse.org/labs}}. The key components of a Zooniverse project are data and workflows. As a sequence of specific tasks, the workflows are tailored by the researchers to gain insights on the data. The workflows are streamlined and optimized balancing the experience of the volunteers with the need to obtain accurate results. 

About 137,000 VERITAS single-telescope images were served on the Muon Hunter website, most of which are preprocessed using the standard two-level cleaning\citep{Cogan08} based on the signal-to-noise ratio of each pixel. A subset of uncleaned images were also uploaded to explore the effect of image cleaning on the results of classification. 
We received a total of $\sim$2.1 million classifications, half within the first week after the official launch of the project, from 5,734 volunteers. 

The workflow for identifying the presence of rings in an image is illustrated in Figure~\ref{fig:flowchart}. We retired an image once a total of 15 volunteers had examined it and finished the workflow. It was possible to draw multiple rings on one image, as sometimes more than one muon was recorded in an image. But this also allowed room for human error, especially when a user was unfamiliar with the workflow. 
After a user was done with the workflow of a subject image, an option to discuss it in the Talk board was presented. The Talk board allows interactions between experts and volunteers regarding specific images. Collections of interesting images are also established by users, including double muon rings and composite images with a muon ring and a shower. 
Moreover, to help new users, a short tutorial, a mini course, as well as a detailed About page were available. 
\begin{figure}[h]
 \centering 
\hspace{-2cm}
 \includegraphics[width=0.38\textwidth]{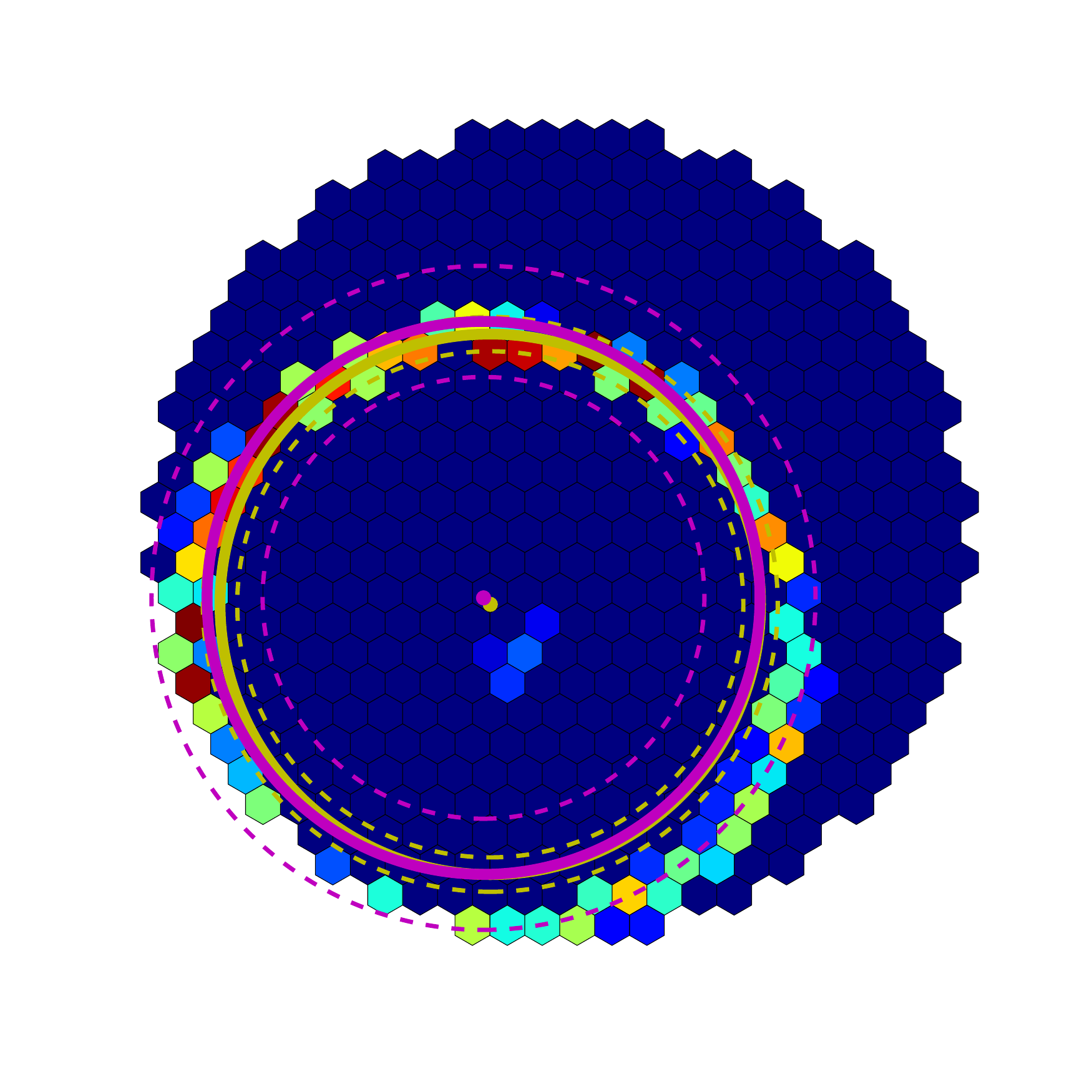} 
\hspace{-0.8cm}
 \includegraphics[width=0.38\textwidth]{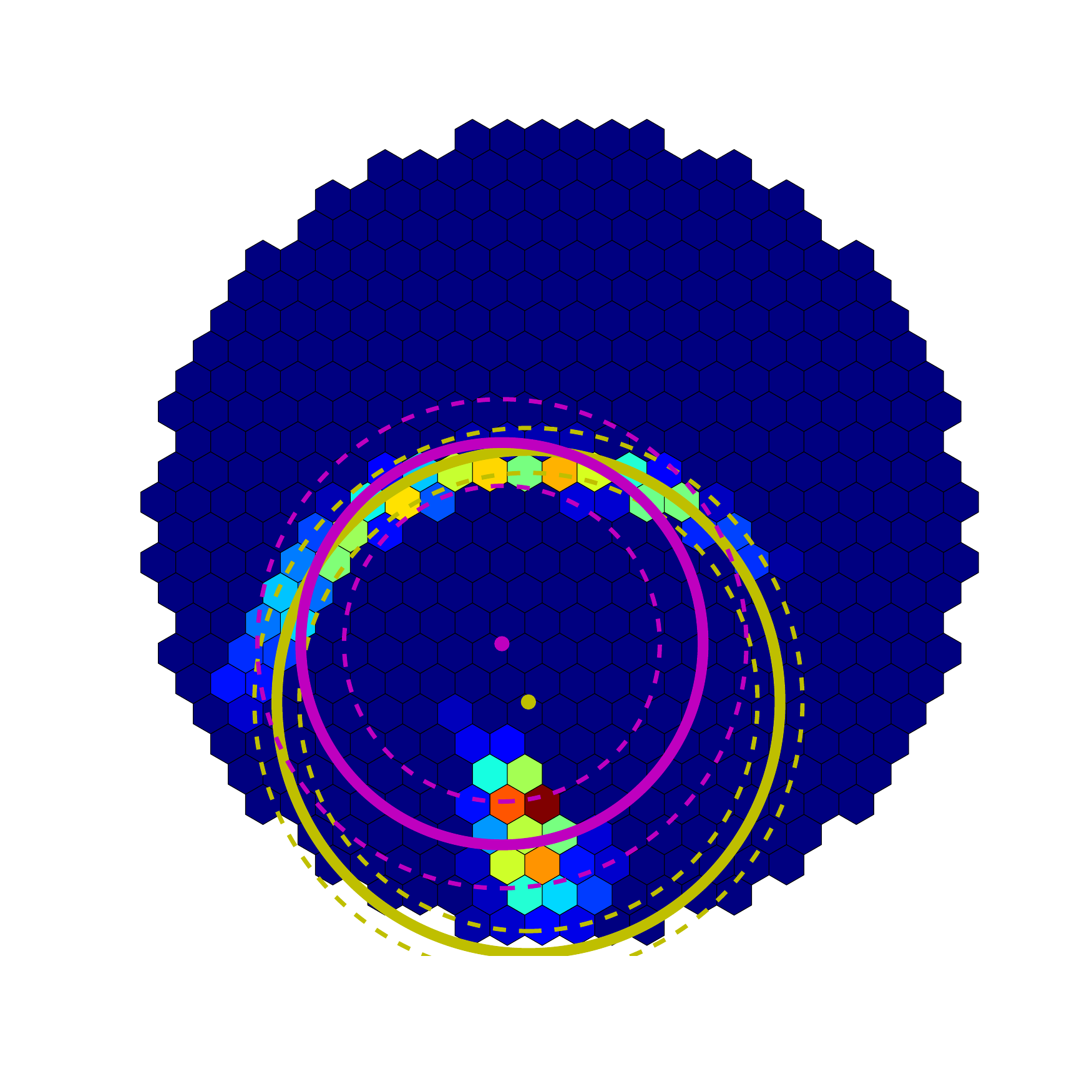} 
\hspace{-0.8cm}
  \includegraphics[width=0.38\textwidth]{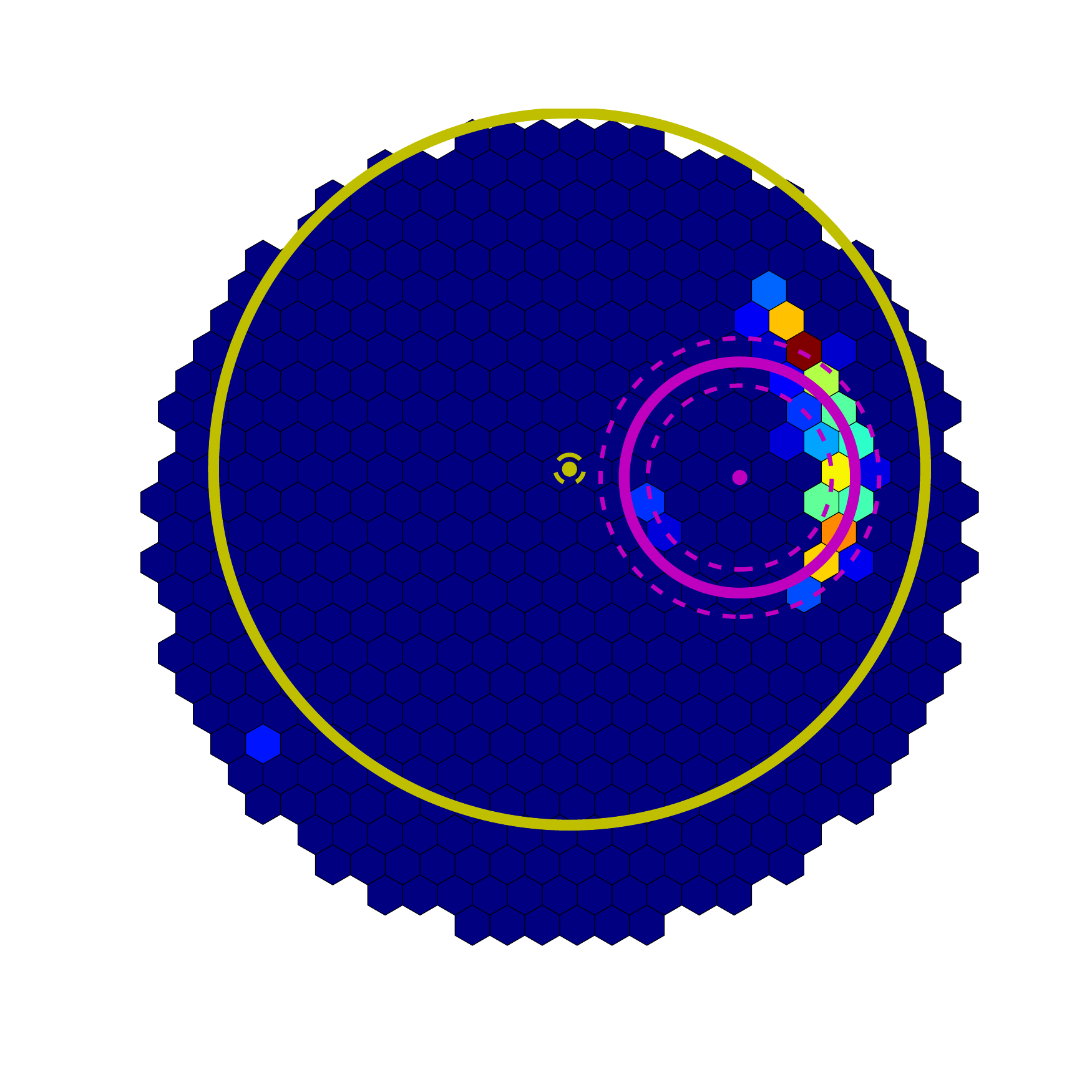} 
  \hspace{-0.8cm} 
\hspace{-0.9cm}
\caption{Examples of VERITAS images with muon rings illustrating the difference between VEGAS analysis and volunteers' opinions. The VEGAS analysis results are shown in magenta, and the Muon Hunter volunteer input is shown in yellow. The circles with solid lines and dashed lines show the mean and the standard deviation of the radius of the ring. 
\label{fig:rings}}
\end{figure}

Figure~\ref{fig:rings} illustrates a few typical comparisons between the ring we obtained from the VEGAS analysis and user input. For a clean image with a single muon ring and no other bright components, as illustrated by the left image, the decision and the best ring drawn by VEGAS and volunteers typically agree. For an image with a muon ring and a CR shower as illustrated by the center image, a human is able to isolate and correctly locate the ring, while VEGAS simultaneously tries to fit the shower component and the actual ring with a single ring. This results in an incorrect ring and may affect subsequent analysis. Sometimes, volunteers behave erratically as illustrated by the right image, and provide incorrect input that can be viewed as outliers, which affects the mean and standard deviation of all user input for a given image. Based on the number of votes, we estimate that roughly 0.8\% of all votes, and 3\% of those votes for the presence of a ring in an image are outliers. 

\begin{figure}[ht!]
 \centering 
\includegraphics[width=0.45\textwidth]{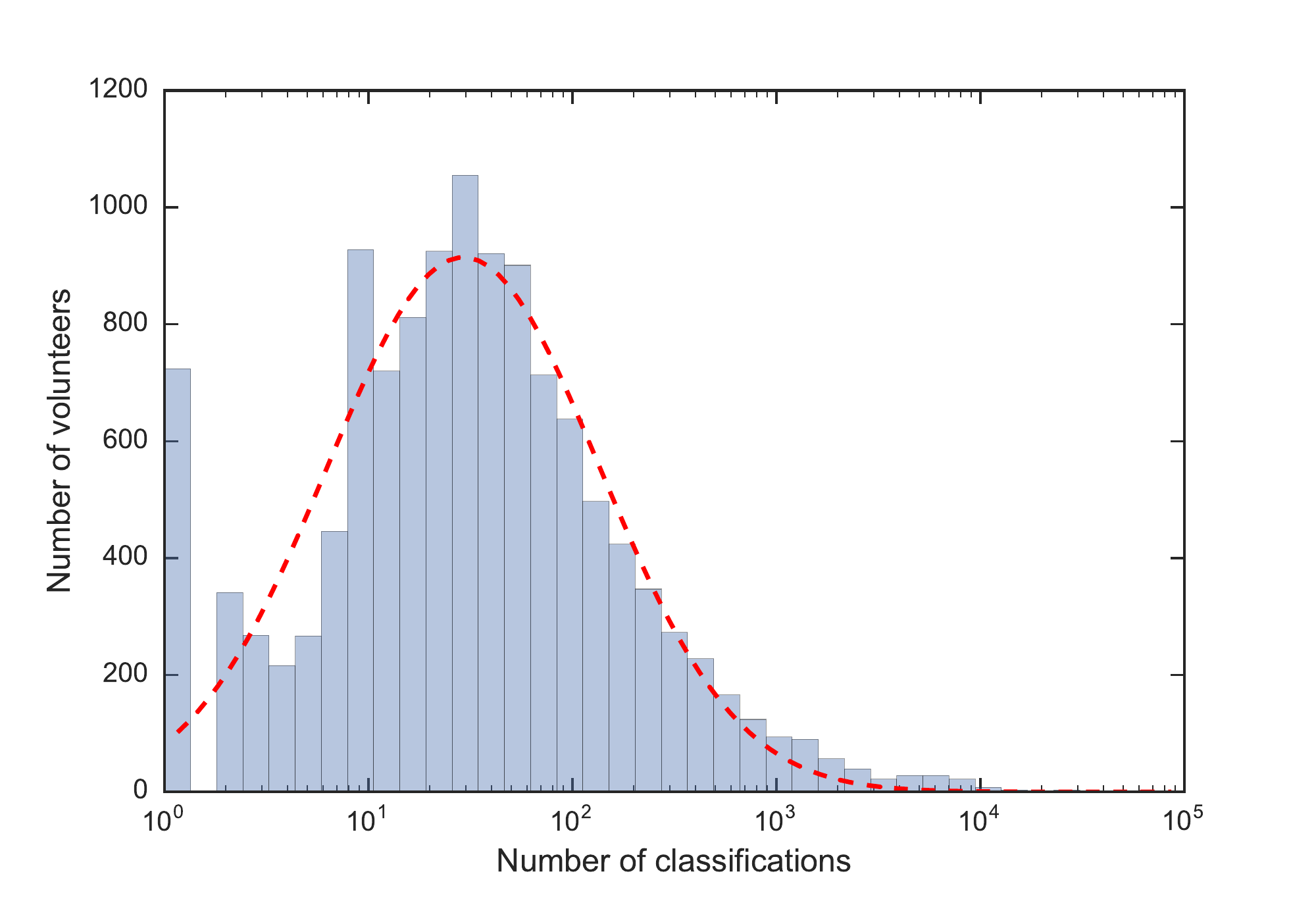}
\includegraphics[width=0.45\textwidth]{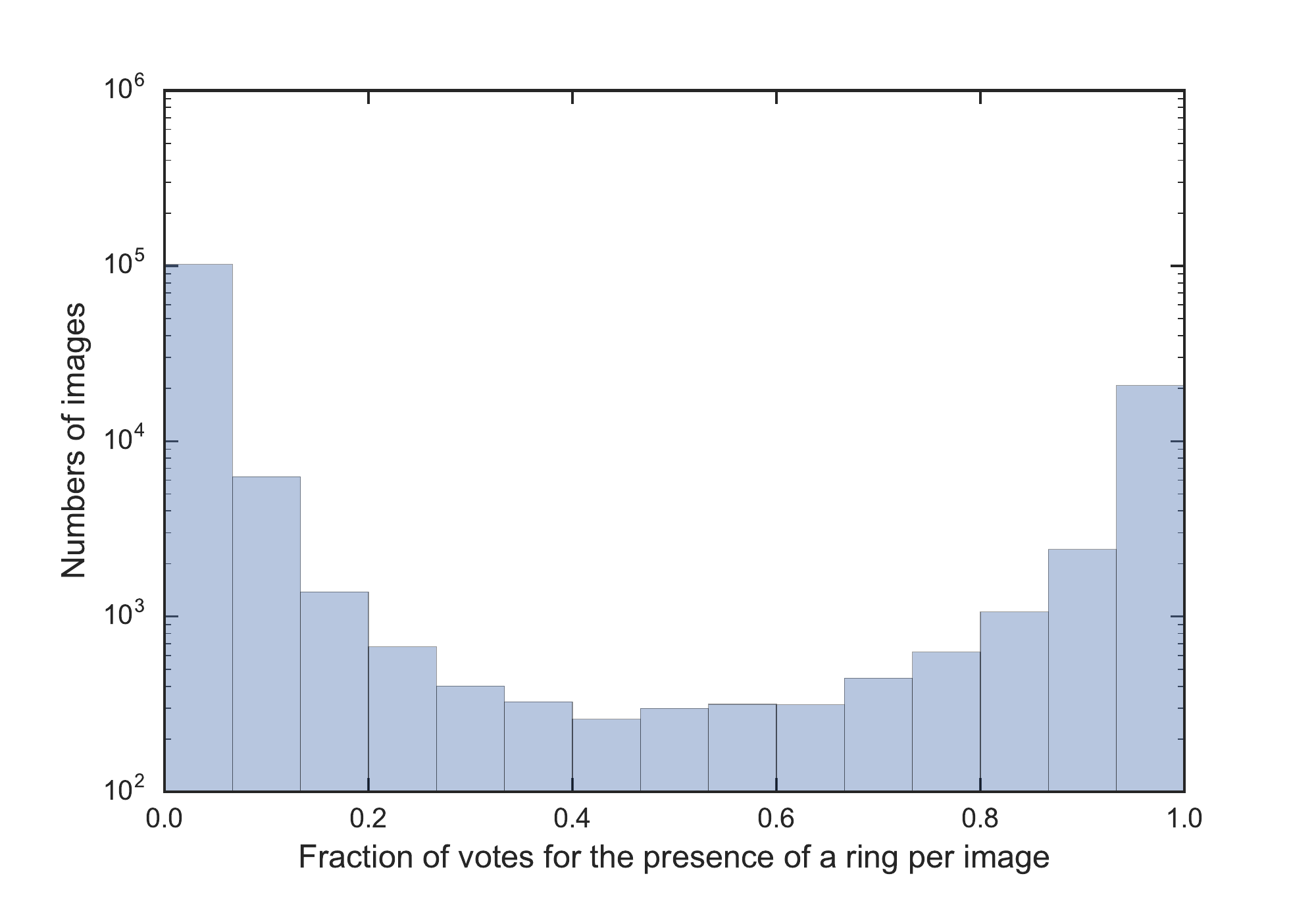}
\caption{Distributions of the input from Muon Hunter users. The left plot shows the distribution of the number of classifications each user made. The right plot shows the distribution of votes for muons each image received. 
\label{fig:user}}
\end{figure}

Figure~\ref{fig:user} summarizes the input we received from the volunteers of Muon Hunter. 
The distribution of the number of classifications each user made roughly follows a log-normal distribution (shown as the red dashed curve), with a median of 30 images per user. 
Assuming the number of classifications from one user is roughly proportional to the time spent, this log-normal distribution is of similar nature to the dwell time of internet users on social media articles \cite{Yin13}. 
There are 16 volunteers who classified more than 10,000 images, while there are 724 volunteers who only classified one image. 

The 15 votes for each image allowed us to estimate the confidence of the users' classifications. 
The distribution of the fraction of votes for the presence of a ring in each image is shown in the right plot of Figure~\ref{fig:user}. 
85\% of the images received unanimous votes from 15 volunteers, 11\% of which were muon events and 73\% of which were non-muon events. 
We chose to label all images with 10 or more votes for muons as muon events, and the rest as non-muon events as a second set of training/test data for building a new CNN model. 

\section{The CNN classification model: training and evaluation}

One purpose of this project is to train a reliable CNN model to classify muon rings. Two sources of labels, provided by the VEGAS analysis and by the Muon Hunter user input, can be used for the training, validation, and testing of the models. A detailed description of the CNN model can be found in \cite{Feng17}, and is summarized below. 

The preprocessing of VERITAS data includes standard image cleaning followed by image oversampling, which we use to approximately convert an image from its original hexagonal coordinates (due to the geometry of the PMT layout) into square coordinates. 
This converts a 499-pixel hexagonal image into a 54 $\times$ 54 pixel square image, and stretches the image by roughly 15\%. 

The oversampled 54 $\times$ 54 pixel images of the muon events and the background events are then used as input features into a CNN model, which is implemented using the \textit{keras} Python deep learning library \citep{Chollet15} with the backend \textit{TensorFlow} \citep{Abadi16}. 
The structure of the CNN model is a simplified ``VGG''-style model \citep{Simonyan14}, with only three layers of small filters, average pooling and dropout in between filter layers, and a two-layer fully-connected neural network classification model after the convolutional layers. 


Treating all images with 10 or more votes for muons 
as muon events, we were able to train a CNN model with a test accuracy of $\sim$97\%, while the best model using VEGAS labels whose test accuracy was $\sim$95\%. We note that the performance of the previous CNN model \cite{Feng17} trained on a smaller set of data worsened when tested with the new, larger dataset, indicating possible overtraining in the previous model.

\section{Summary and outlook}
\label{sec:sum}

We received a phenomenal response from volunteers to the Muon Hunter project. The input from volunteers helped us gain insight into where the standard analysis is lacking, and train an updated machine learning model using convolutional neural networks. 

We are working on isolating images with only one muon ring without other components (e.g. cosmic-ray shower) based on the input from users, and using them as a set of less noisy training examples to further improve the performance of the classification model. The single-muon images only have one relevant radius value, and therefore allow us to train a regression model to predict the radius of the muon ring. 
More questions will be designed for a future workflow to augment this purpose (e.g. whether a muon image also contains a cosmic ray shower image). 
Different criteria for converting the user input into a set of labels for the training of the CNN model are also being explored. By introducing a more strict cut on the number of users agreeing on a given image being muon or non-muon, a less ambiguous but smaller data set can be used.

\acknowledgments
VERITAS is supported by grants from the U.S. Department of Energy Office of Science, the U.S. National Science Foundation and the Smithsonian Institution, and by NSERC in Canada. We acknowledge the excellent work of the technical support staff at the Fred Lawrence Whipple Observatory and at the collaborating institutions in the construction and operation of the instrument. 
The VERITAS Collaboration is grateful to Trevor Weekes for his seminal contributions and leadership in the field of VHE gamma-ray astrophysics, which made this study possible.

The authors also gratefully acknowledge all the Muon Hunter volunteers who contributed to this effort without whom this work would not be possible. 
Muon Hunters was developed with the help of the ASTERICS Horizon2020 project. ASTERICS is a project supported by the European Commission Framework Programme Horizon 2020 Research and Innovation action under grant agreement n. 653477. 

\vspace{5mm}

\printbibliography

\end{document}